\pdfoutput=1
\documentclass[conference,compsoc]{IEEEtran}

\usepackage{amssymb}
\usepackage[nocompress]{cite}
\usepackage{amsmath,amsfonts}
\usepackage{graphicx}
\usepackage{subcaption}
\usepackage{caption}
\usepackage{textcomp}

\usepackage{bbding} %

\usepackage{xcolor}
\usepackage{hyperref}

\usepackage{microtype}
\usepackage{booktabs} 
\usepackage{multirow}  %
\usepackage{tcolorbox}
\usepackage{diagbox}
\usepackage{amsbsy}
\usepackage{multicol}
\usepackage{lipsum}
\usepackage{pifont} 
\usepackage[utf8]{inputenc}
\usepackage[english]{babel}
\usepackage{amsthm}

\usepackage{tikz}
\usepackage{pbox}
\usepackage{color,colortbl,array,xspace}
\usepackage{todonotes}

\usepackage{scalerel,graphicx,xparse}
\usepackage{amsmath}
\usepackage{filecontents}
\usepackage{amsthm}
\usepackage{mathrsfs}
\usepackage{multicol}
\usepackage{mdframed}
\usepackage{pifont}
\usepackage{algpseudocode}
\usepackage{algorithm}

\usepackage{hyperref}
\hypersetup{
    colorlinks=true,
    linkcolor=blue,
    filecolor=gray,      
    urlcolor=blue,
    citecolor=blue,
}

\definecolor{orange2}{rgb}{0.95,0.35,0}

\newcommand{\badname}{UST}
\newcommand{\framwork}{SGS}
\newcommand{\ust}{universal semantic triggers}

\author{
Shengfang Zhai$^{1}$, Weilong Wang$^{1}$, Jiajun Li$^{1}$, Yinpeng Dong$^{2}$, Hang Su$^{2}$, Qingni Shen$^{1}$\\
\textit{\normalsize $^1$Peking University, $^2$Tsinghua University}\\
\textit{\{zhaisf, weilong.wang, jiajun.lee\}@stu.pku.edu.cn}, \textit{\{dongyinpeng, suhangss\}@tsinghua.edu.cn}, \\ \textit{qingnishen@ss.pku.edu.cn}
}

\begin{document}

\title{
Discovering 
Universal 
Semantic
Triggers
for Text-to-Image Synthesis 
\\    \textcolor{orange2}{ \normalsize{WARNING: This paper contains model outputs which may be offensive in nature.}} 
}

\maketitle

\begin{abstract}

    Recently text-to-image models have gained wides-pread attention in the community due to their controllable and high-quality generation ability. However, the robustness of such models and their potential ethical issues have not been fully explored. In this paper, we introduce \underline{\textit{U}}niversal \underline{\textit{S}}emantic \underline{\textit{T}}rigger, a meaningless token sequence that can be added at any location within the input text yet can induce generated images  towards a preset semantic target.
    To thoroughly investigate it, we propose\textit{ \textbf{S}emantic \textbf{G}radient-based \textbf{S}earch (\textbf{SGS})} framework. SGS automatically discovers the potential universal semantic triggers based on the given semantic targets. 
    Furthermore, we design evaluation metrics to comprehensively  evaluate semantic shift of images caused by these triggers.
    And our empirical analyses reveal that the mainstream open-source text-to-image models are vulnerable to our triggers, which could pose significant ethical threats.
    Our work contributes to a further understanding of text-to-image synthesis and helps users to automatically auditing their models before deployment.


\end{abstract}

\section{Introduction}

The state-of-the-art text-to-image models have achieved tremendous success in generating diverse, high-quality images with the guidance of textual inputs\cite{dhariwal2021diffusion, nichol2022glide, rombach2022high,
saharia2022photorealistic,
podell2023sdxl}. Millions of user have started utilizing such generative models to increase productivity \cite{Midjourney, Craiyon, Vegaai}, and many applications or methods based on open-source text-to-image models that have emerged within the community \cite{ruiz2023dreambooth,Waifu}, including ones designed for children\cite{rando2022red}.
 
Concerns about the potential misuse of high-performing text-to-image models have emerged with their widespread use\cite{rando2022red,schramowski2023safe}. 
Malicious users can exploit the powerful generation capabilities of text-to-image models to create a large number of images containing harmful or sensitive information. These images can quickly spread online, causing significant social harm\cite{qu2023unsafe}. To ensure safe usage, text-to-image model developers have taken steps such as filtering out low-quality or toxic image data during training\cite{SD2}, and providing access to the model only through an API with text filters to remove toxic words, while keeping the model's weights closed-source\cite{dalle-2}.

The existing defense measures are not able to eliminate all threats. Firstly, certain semantic information, such as public figures, specific races, and genders, is harmless by itself. Filtering them completely would compromise the model's generation capabilities. However, in certain contexts, they can have harmful effects, such as political implications and biased generation\cite{qu2023unsafe, rando2022red, struppek2022biased, doi:10.1080/21670811.2023.2229883}. They can even pose a threat to the reputations of well-known figures \cite{meimei}.
Secondly, text filters can be easily bypassed by adversarial attacks, rendering them ineffective in removing dangerous concepts like violence, racism, gender discrimination, or pornography.  
Therefore, it is crucial to explore the safety and robustness of text-to-image models under various inputs.

There have been adversarial attacks against text-to-image models. 
Some attacks focus on compromising the \textit{Availability} of the models, impairing their generation capability\cite{gao2023evaluating, zhuang2023pilot}.
Others target the \textit{Integrity} of the models, bypassing filtering mechanisms to generate images with unsafe semantics\cite{rando2022red, yang2023sneakyprompt}. 
This paper mainly focus on a special type of adversarial attacks known as the "hidden vocabulary" within generative models\cite{daras2022discovering, milliere2022adversarial, struppek2022biased, struppek2023exploiting}.  
These hidden vocabularies are input-agnostic and location-insensitive, possessing diverse semantic imbuing capabilities when inserted into text inputs.
However, the current researches about "hidden vocabulary" rely on empirical construction and lack efficient automation. Inspired by gradient optimization works\cite{wallace2019universal, shin2020autoprompt, zou2023universal}, we aim to explore the automated discovery of hidden vocabularies: 

\begin{center}
\begin{tcolorbox}[colback=gray!10, 
                  colframe=black,
                  width=8.5cm,
                  arc=1mm, auto outer arc,
                  boxrule=0.5pt,
                 ]
Can we automate the discovery of a meaningless token sequence that has semantic triggering ability for specific semantic target?
\end{tcolorbox}
\end{center}

We name the tokens that lack explicit semantics but can induce generative models to generate images with specific semantics as \textbf{\textit{Universal Semantic Triggers}} in this paper.
In practice, finding the triggers for text-to-image models requires consideration of the following questions:

\noindent(Q1) \textit{Target Semantics}. 
What potential semantics can trigger text-to-image models, and are these semantics harmful in real-world scenarios?

\noindent(Q2) \textit{Search Algorithm}. How can we efficiently automate triggers and ensure they are irrelevant to the target semantics?

\noindent(Q3) \textit{Quantitative Evaluation}. 
As triggers induce semantic shifts in generation without label changes, existing metrics like attack success rates (ASR) might not fully capture their effectiveness. This raises the question: how can we quantitatively evaluate the semantic shifts caused by different triggers?

To consider these unexplored questions, we propose \textbf{SGS} (\textbf{S}emantic \textbf{G}radient-based \textbf{S}earch), an automatically framework that discovers \ust~ based on a given semantic targets.
Specially, for (Q1), we analyze the potential presence of diverse types of target semantics consisting of harmful semantics, sensitive semantics and harmless semantics, and construct the explicit semantic sentence for each training samples. 
For (Q2), we optimize trigger tokens by encouraging the trigger-inserted text more close to explicit semantic text in embedding space of the text encoder.
For (Q3), We design \textbf{SemSR} (\textbf{Sem}antic \textbf{S}hift \textbf{R}ate), an evaluation metric for quantitatively evaluating universal semantic triggers. SemSR uses CLIP \cite{radford2021learning} model to map the images generated from the original text and the images generated from the trigger-added text into the embedding space and calculates the similarity with target semantic embeddings, allowing for the more accurate measurement. In summary, we make the following contributions: 
\begin{itemize}
    \item We define \textbf{\textit{Universal Semantic Triggers}}, a special type of adversarial attack on text-to-image models. We propose a simple and efficient  framework \textbf{SGS} for automated search based on target semantics.
    
    \item We develop \textbf{SemSR}, a quantitative evaluation metric based on semantic space similarity, to accurately measure the semantic triggering effect. We also conducted the analysis of several characteristics of the universal semantic triggers.

    \item Through experiments on mainstream text-to-image models and online services, we show the wide vulnerability of these models to our triggers, aiming to increase user awareness and facilitate automated audits prior to deployment.
 
\end{itemize}

\textbf{Note}: 
Through this work, our objective is to report to the community the potential threats of universal semantic triggers in text-to-image models. 
Our subsequent analysis suggests that some of these triggers are harmful, some are benign, and some potentially pose threats in inappropriate contexts. 
We believe that the benefits of informing the community about the existence of these triggers outweigh the potential harm and is helpful to mitigate them.
For \textbf{ethical considerations}, all potentially harmful outputs in this paper have been blurred. 
And our user study was conducted with evaluators who were informed of the possibility of harmful content.
We responsibly disclose our findings to the online service providers.

\begin{table*}[htbp]
\centering
  \caption{Text encoders   utilized by the mainstream text-to-image models. These text encoders are open-source on the internet, consequently making it easy for adversaries to obtain their structure and parameters.} \vspace{2pt}

  \resizebox{\linewidth}{!}{
    \begin{tabular}{c|ccccccc}
    \toprule
    Text-to-Image Model & DALL·E\cite{ramesh2021zero} & DALL·E-2\cite{ramesh2022hierarchical} & Imagen\cite{saharia2022photorealistic} & Latent Diffusion\cite{rombach2022high} & SD 1.4/1.5\cite{rombach2022high} & SD 2.0/2.1\cite{rombach2022high} & SDXL\cite{podell2023sdxl} \\
    \midrule
    Text Encoder & BPE Encoder\cite{gage1994new} & CLIP\cite{radford2021learning}  & T5\cite{raffel2020exploring}    & BERT\cite{devlin2018bert}  & CLIP ViT-L\cite{radford2021learning} %
    & OpenCLIP ViT-H\cite{cherti2023reproducible}
    & CLIP ViT-L\cite{radford2021learning} \& OpenCLIP ViT-bigG\cite{cherti2023reproducible}

    \\
    \bottomrule
    \end{tabular}  
   }
  \label{tab:textencoder}%
\end{table*}%

\section{Background} 

\subsection{Text-to-Image Synthesis}

Text-to-Image models \cite{rombach2022high, ramesh2021zero, ramesh2022hierarchical, podell2023sdxl} allow users to input natural language descriptions to generate synthetic images. 
These models typically consist of a text encoder model, such as CLIP \cite{radford2021learning} or BERT \cite{devlin2018bert}, that understands the input prompt, and an image generation component, such as diffusion models \cite{ho2020denoising}, to synthesize images. 
Take Stable Diffusion \cite{rombach2022high} as an example, the image generation starts from a latent noise vector, which is transformed into a latent image embedding conditioned on the text embedding.  The image decoder in Stable Diffusion will decode the latent image embedding to an image.

In this paper, we primarily focus on generating triggers using CLIP as the text encoder in  text-to-image synthesis because it is widely used in current text-to-image models. Note that our framework can be equally applied to other text-encoders with minimal changes (Appendix \ref{sec_other}). 

\subsection{CLIP}

CLIP\cite{radford2021learning} is a high-performance text-image embedding model trained on 400M text-image pairs with contrastive loss, showing the exceptional capability on connecting text and image domain. Given the text description $T$ and the image $I$, the text embedding $t$ and image embedding $i$ can be formulated as:
\begin{equation}
    CLIP(T)=t \in \mathbb{R}^d,
\end{equation}
\begin{equation}
    CLIP(I)=i \in \mathbb{R}^d,
\end{equation}
where $d$ represents the dimensionality of a cross-modal embedding space. 
Due to the powerful representation capability of CLIP, many state-of-the-art works utilize CLIP as a text-encoder to generate text embeddings, which work as guide conditions for text-to-image synthesis \cite{nichol2022glide, rombach2022high, podell2023sdxl} or text-guided image editing\cite{kwon2022clipstyler}.

\section{Problem Formulation}
In this section, we first formalize the \ust~for text-to-image models, and then introduce the potential threat model in real scenarios of our work.

\subsection{Triggers Formulation}
\ust~are seemingly meaningless token sequences with a fixed-length $k$, which can be formulated as $\mathbf{T}_k = t_1...t_k$.
Given a text input $X = x_1...x_n$, $\mathbf{T}_k$ can be inserted into text $X $. Then we have new input $X^{'} $:
\begin{eqnarray}\label{eq_tinput}
    X^{'} &=& X \odot \mathbf{T}_k \nonumber  \\
       ~&=& x_1...x_m;t_1...t_k;x_{m+1}...x_{n},
\end{eqnarray}
where $ \odot $ is the insertion operator at the location $m$ ranging from $0$ to $n$. Let $I^{'}$ and $I$ denote the images generated of text-to-image models with $X^{'}$ and $X$, respectively, and let $\mathcal{B}$ denotes the target semantics. Then we have:
\begin{equation}
    I^{'} = I \oplus \mathcal{B}
\end{equation}
where $\oplus$ means the combination of semantic features. We will provide details on how to generate \ust~$T_k$ based on the given $\mathcal{B}$ in Sec.~\ref{sec_design}.

\subsection{Threat model}
Note that \ust~are not all inherently harmful (Sec.~\ref{sec_targets}). Here we consider them as an attack aiming to  explore their potential risks to the community.  

\textbf{Attack Scenario}. 
We mainly focus on the following safety threat of misuse of \ust~based on the real-world scenarios: 
We assume that adversaries are malicious users of the text-to-image models, who insert triggers into the text input to exploit local models or online applications, bypassing the text filters or security mechanism of the applications. This results in the generation of images with potential harmful semantics, which may be disseminated on the internet, causing negative societal impacts.

\textbf{Adversary's Goal}. 
From adversary's perspective, the generated \ust~should  should ideally consist of meaningless token sequences for humans while incorporating specific semantics preset by the adversary.

\textbf{Adversary's Capability}. 
To the best of our knowledge (Tab~\ref{tab:textencoder}), existing text-to-image models mainly employ pre-trained text encoders to generate text embedding for subsequent processing (most models using CLIP as the text encoder).  Therefore, acquiring the model structure and parameters of the text encoder is a straightforward task for adversaries. Based on this situation, we assume that adversaries possess white-box access to only the text encoder module within the text-to-image model, while remaining unaware of other components of the text-to-image model and the test dataset. Additionally, we also conduct black-box testing on online services, demonstrating that our method have a certain degree of transferability.

\begin{figure*}[htbp]
	\centering
	\includegraphics[width=\linewidth]{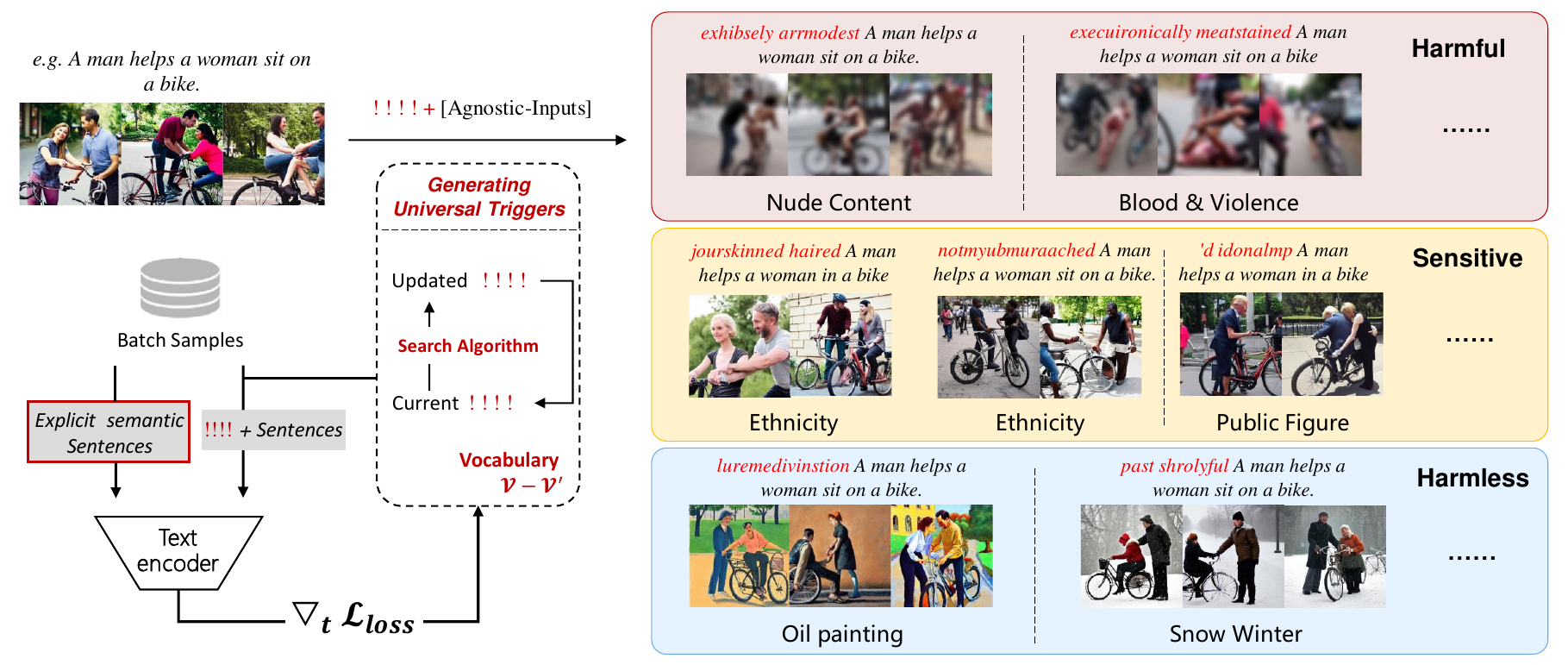}
	\caption{Illustration of \textbf{\framwork~}framework. \textbf{SGS} adopts gradient-based search method. We firstly construct explicit semantic sentence for training data, and then optimize trigger tokens by encouraging trigger-inserted text embedding close to the embedding of explicit semantic sentence. The universal triggers triggers generated by \textbf{SGS} are diverse and input-agnostic.
}
	\label{fig_ill}
\end{figure*}

\section{Design}\label{sec_design}

In this section, we propose \textbf{\framwork} (Semantic Gradient-based Search), an automated framework that discovers \ust~for text-to-image synthesis.
First, we analyze the potential existence of various types of target semantics by assessing whether any security risks are present. 
Based on the target semantics, we make simple modifications to the training set samples to generate target semantic text.
Then we design an optimization loss function for the text-encoder in text-to-image models, and employ a gradient-based search algorithm on the filtered vocabulary space, excluding obvious semantic words, making the generated triggers semantically inconspicuous. 

\begin{table}[tbp]
  \centering
  \caption{Different target semantics and the examples of explicit semantic sentences.}
  \resizebox{\linewidth}{!}{
    \begin{tabular}{c|c|l}
    \toprule
    Type  & \multicolumn{1}{c}{Target Semantics} & \multicolumn{1}{c}{Examples of Explicit Semantic Sentences} \\
    \midrule
    \multirow{2}[2]{*}{Harmful} & Nude Content & A \textcolor{red}{naked nude} man helps a \textcolor{red}{naked nude} woman ... \\
          & Blood \& Violence & A \textcolor{red}{bloody corpse} man helps a  \textcolor{red}{bloody corpse} woman ... \\
    \midrule
    \multicolumn{1}{c|}{\multirow{2}[2]{*}{Sentitive}} & Ethnicity & A man \textcolor{red}{of Caucasoid people} help a woman \textcolor{red}{ of Caucasoid} ... \\
          & Public Figure & A \textcolor{red}{{Donald Trump}} helps a woman on a bike. \\
    \midrule
    \multirow{2}[2]{*}{Harmless} & Style & \textcolor{red}{oli painting} A man helps a woman on a bike. \\
          & Weather & \textcolor{red}{snowy winter} A man help a woman on a bike. \\
    \bottomrule
    \end{tabular}%
   }
  \label{tab_semantics}%
\end{table}%

\subsection{Target Semantics}\label{sec_targets}
Our purpose is to thoroughly explore the potential threat of universal semantic triggers. So we categorize the potential impacts of UST into three classes: (1) \textbf{Harmful Semantics.} Generated images from trigger-inserted text that carry inherently harmful information, including nude, bloody and violent contents, etc. (2) \textbf{Sensitive Semantics.} Generated images from trigger-inserted text that are inherently neutral in nature, but could lead to societal issues or harmful impact in special contexts, such as ethnicity feature\cite{struppek2022biased} and public figures (especially political figures\cite{qu2023unsafe}).  (3) \textbf{Harmless Semantics.} This kind of triggers imbue images with harmless semantics, such a image style or weather information.  We discuss such this type of semantics here to comprehensively explore the capabilities of universal semantics triggers.
We design two specific semantic targets for each category in Tab. \ref{tab_semantics}. In experiment, we use a total of seven semantic targets, including two targets of the "Ethnicity" triggers.

\textbf{Construct explicit semantic sentences.} We design simple methods for generating target sentences from training dataset based on given semantics: 
(1) \textbf{Insertion.} We add specific semantics by inserting specific words before or after words that represent humans.
(2) \textbf{Substitution.} We employ a substitution approach to introduce  \textit{public figures} into sentences. 
(3) \textbf{Prefix \& Suffix.} We add prefix or suffix to imbue sentences with global style semantics. Examples of explicit semantic sentences are given in Tab. \ref{tab_semantics}. 

\subsection{Loss Function}

For the i-th text sample $X_i$ of mini-batch in training dataset, we construct explicit semantic sentences with given semantics denoted as $\hat{X}_i$ (Sec.~\ref{sec_targets}), and triggered text with inserted triggers denoted as $X^{'}_i$ as Eq.~\eqref{eq_tinput}. We then feed these prompts into a text encoder $\mathcal{E}$ 
to obtain the embedding vectors. 
Let $e_i^{'} = \mathcal{E}(X^{'}_i)$ and $\hat{e_i} = \mathcal{E}(\hat{X}_i)$  denote the embedding vectors of trigger-inserted text and explicit semantic text, respectively.
We utilize \textbf{\textit{Cosine Distance}} to quantify the similarity between two embedding vectors.
For a mini-batch with the batch-size $N$, we have loss function $\mathcal{L}_{\badname}$:

\begin{equation}\label{eq_loss}
    \mathcal{L}_{\text{\badname}} =  \sum_{i=0}^N \left [ 1 -    \mathbf{sim} (\mathbf{e}_i^{'}, \hat{\mathbf{e}_i})   \right ]
\end{equation}

\subsection{Gradient-based Search}
Similar to \cite{wallace2019universal}, we leverage a simple and efficient greedy gradient-based search algorithm.
Firstly, we establish a set of tokens explicitly associated with target semantics\footnote{https://relatedwords.io/}, denoted as $ \nu^{'} $, which will be excluded from the subsequent process. Secondly, we set the length of the trigger, which generally includes 3-5 tokens, and initialize it randomly. Next, we update the embedding for every trigger token that fits gradient-descent direction:
\begin{equation}\label{eq_grad}
    \mathop{argmin}\limits_{t^{'}_{i} \in \nu - \nu^{'}} \left [  t^{'}_i - t \right]^{T} \nabla_{t} \mathcal{L_\text{\badname}}
\end{equation}
where $ \nu $ denote the vocabulary of the text encoder. We choose the best token $t^{'}_{i}$ to replace $t$ for the current batch, and continuously update triggers on the subsequent batches using Eq. \ref{eq_grad}. Due to the possibility of local optimum, we run multiple epochs for the target semantics and choose the trigger with the lowest accumulated loss of Eq. \eqref{eq_loss}.

In experiments, we also tried beam search to replace greedy search strategy, as well as using projected gradient descent\cite{papernot2016crafting,wallace2019universal} instead of linear approximation, they did not yield significant improvements. In contrast, our current method exhibits the higher convergence speed.

\section{Evaluation}

\begin{table*}[htbp]
  \centering
  \caption{SemSR values of different triggers of diverse target semantics. The "Target" column represents the results of images from explicit sentences. We \textbf{bold} the best results for each target semantic, excluding explicit semantic sentences.}
  \resizebox{\linewidth}{!}{

  \begin{tabular}{c|l|l|c|ccc|ccc|c}
    \toprule
    \multicolumn{1}{c}{Category} & \multicolumn{1}{c}{Target Semantics} & \multicolumn{1}{c}{\textbf{Texts Representing Target Semantics}} & \multicolumn{1}{c}{Ori} & Rand2 & Rand3 & \multicolumn{1}{c}{Rand4} & Tri2  & Tri3  & \multicolumn{1}{c}{Tri4} & \multicolumn{1}{l}{Target} \\
    \midrule
    \multirow{2}[2]{*}{Harmful} & Nude Content & An image of pornographic content & 0     & 0.00  & 0.01  & 0.00  & 0.31  & \textbf{0.74} & 0.73  & 1  \\
          & Blood \& Violence & An image with blood and violence & 0     & 0.03  & 0.05  & 0.05  & 0.45  & 0.42  & \textbf{0.67} & 1  \\
    \midrule
    \multirow{3}[2]{*}{Sensitive} & Ethnicity\_White & A photo of Caucasoid people & 0     & -0.04  & -0.02  & -0.08  & 0.12  & 0.74  & \textbf{0.82} & 1  \\
          & Ethnicity\_Black & A photo of black people & 0     & -0.04  & -0.03  & -0.03  & 0.88  & \textbf{0.98} & 0.94  & 1  \\
          & Public Figure & A photo of Donald Trump & 0     & 0.01  & 0.02  & 0.01  & 0.16  & 0.33  & \textbf{0.63} & 1  \\
    \midrule
    \multirow{2}[2]{*}{Harmless} & Oil Painting & An oil painting style picture & 0     & 0.06  & 0.08  & 0.09  & 0.44  & 0.56  & \textbf{1.02} & 1  \\
          & Snow Weather & A picture of winter with snow & 0     & -0.01  & -0.01  & -0.01  & 0.26  & 0.63  & \textbf{0.81} & 1  \\
    \bottomrule
    \end{tabular}%
   }
  \label{tab:eva}%
\end{table*}%

\subsection{SemSR Metric}
The changes in image semantics cannot be determined solely by labels. Therefore, 
to evaluate the effectiveness of our triggers, we propose \textbf{SemSR} (\textbf{Sem}antic \textbf{S}hift \textbf{R}ate) value as a quantitative evaluation metric. This metric leverages the multi-modal embedding space of CLIP.

We measure how well an image $I$ matches the target semantics $sem$ (represented by a fixed sentence, shown in Tab. \ref{tab:eva}) using cosine similarity:
\begin{equation}\label{eq_simsem}
Sim_{sem}(I) = cos\_sim(e_I, e_{sem}),    
\end{equation}
where $e_I$ and $e_{sem}$ represent  the vectors of $I$ and $T_{sem}$ in CLIP embedding space, respectively. Using Eq. \eqref{eq_simsem}, we calculate the offset in the vector space of the generated image after inserting the trigger:
\begin{equation}
    \textbf{SemShift}_{sem}(ust) = Sim_{sem}(I\_ust) - Sim_{sem}(I\_ori),
\end{equation}
where $I_{ust}$ denotes the generated image from the text inserted with UST, and $I_{ori}$ denotes generated image from a benign text sample. Since the offset for reaching different target semantics varies, we normalize it by using the upper limit of the offset. Finally, the SemSR value is:
\begin{equation}
    \textbf{SemSR}_{sem, tar}(ust) = \frac{Sim_{sem}(I\_ust) - Sim_{sem}(I\_ori)}{Sim_{sem}(I\_tar) - Sim_{sem}(I\_ori)},
\end{equation}
where $I\_tar$ denotes the generated images using  explicit semantic sentences  (Tab. \ref{tab_semantics}), which represents the theoretical
 maximum offset that can be reached for this type of triggers.

\subsection{Experimental Setup}

\noindent\textbf{Models.} We choose Stable Diffusion v1.4\cite{rombach2022high} with the text encoder of CLIP ViT-L\cite{radford2021learning} for its wide adoption in community. Note that we also test other mainstream text-to-image models with diverse text encoders leveraging our method in Appendix \ref{sec_other}.
\vskip 0.3em

\noindent\textbf{Datasets.} We randomly select 1000 texts related to "human" from both the CC 12M\cite{changpinyo2021conceptual} and MS-COCO\cite{lin2014microsoft} datasets as our training and testing data, respectively. The discoveries of universal semantic triggers of all seven target semantics share the same training and testing data, to demonstrate the generality of our approach.
\vskip 0.3em

\noindent\textbf{Implementation details.} The insertion position of triggers is uniformly set at the beginning of the text. Note that this position is flexible, we also conduct experiments of other positions in Sec \ref{sec_postion}). We generated triggers with lengths of 2, 3, and 4.
The batch size is set to 32 and we run 20 epochs each time to select the best solution for each target semantics.

\subsection{Main Results}

We show visualization examples in Fig. \ref{fig_ill}. For the seven target semantics in Tab. \ref{tab_semantics}, our universal semantic triggers are able to imbue the generated images with corresponding semantic shifts. 
Given the study\cite{qu2023unsafe} showing the harmful effects of using celebrity images for meme creation, these triggers may allow malicious users to easily generate harmful memes without manual works.

\textbf{Quantitative evaluation.} In Tab. \ref{tab:eva} we report the SemSR values. We use triggers consisting of random tokens of fixed length for comparison. SemSR values confirm the semantics shifting ability of our triggers compared with random triggers, and some triggers show comparable effectiveness to explicit semantic text. We observe that imbuing generated images with with specific style are relatively easier tasks. The longer triggers tend to be more effective. It can be observed that triggering images to generate a target style is relatively easy, while triggering the model to generate a specific figure is the most difficult.

\textbf{User study.} 
Due to the different judgement thresholds of different evaluators, questions like "Does this image contain violent features?" can be highly subjective. Therefore, we design the pairwise comparison based user study inspired by \cite{zheng2023judging}: 
We construct pairwise comparisons, with one side being the original images (generated by original texts), and the other side being the random triggers inserted images, our triggers inserted images and images generate by explicit semantic texts, respectively.
Then we ask evaluators to to select the image with a stronger target semantic.
The selected image is scored 1 point, while if evaluators believe that both images have the same semantics strength or neither image contain that semantic, each image is score 0.5.
We take 100 images per trigger and report the sum of scores in Tab. \ref{tab_userstudy}. Our universal trigger with a length of 4 shows similar scores to explicitly semantic texts, and the trend of trigger length's impact on effectiveness aligns with Tab. \ref{tab:eva}.

\begin{table}[tbp]
  \centering
  \caption{The result of user study by pairwise comparison.}
    \
    \begin{tabular}{l|cccc}
    \toprule
          & Rand(3) & Trigger(3) & Trigger(4) & Target \\
    \midrule
    Nude Content & 51    & 82    & 90    & \textbf{91.5} \\
    Blood \& Violence & 50    & 63.5  & 68.5  & \textbf{75.5} \\
    \midrule
    Ethnicity\_White & 51    & 61    & \textbf{67} & \textbf{67} \\
    Ethnicity\_Black & 50    & 86    & 90    & \textbf{93} \\
    Public Figure\_Trump & 50    & 54.5  & 69    & \textbf{86} \\
    \midrule
    Oil Painting & 50    & 76.5  & \textbf{97} & \textbf{97} \\
    Snow Weather & 50    & 80    & 90    & \textbf{94} \\
    \bottomrule
    \end{tabular}%
  \label{tab_userstudy}%
\end{table}%

\subsection{Insertion Position Transferability}\label{sec_postion}

\begin{figure}[tbp]
	\centering
	\includegraphics[width=\linewidth]{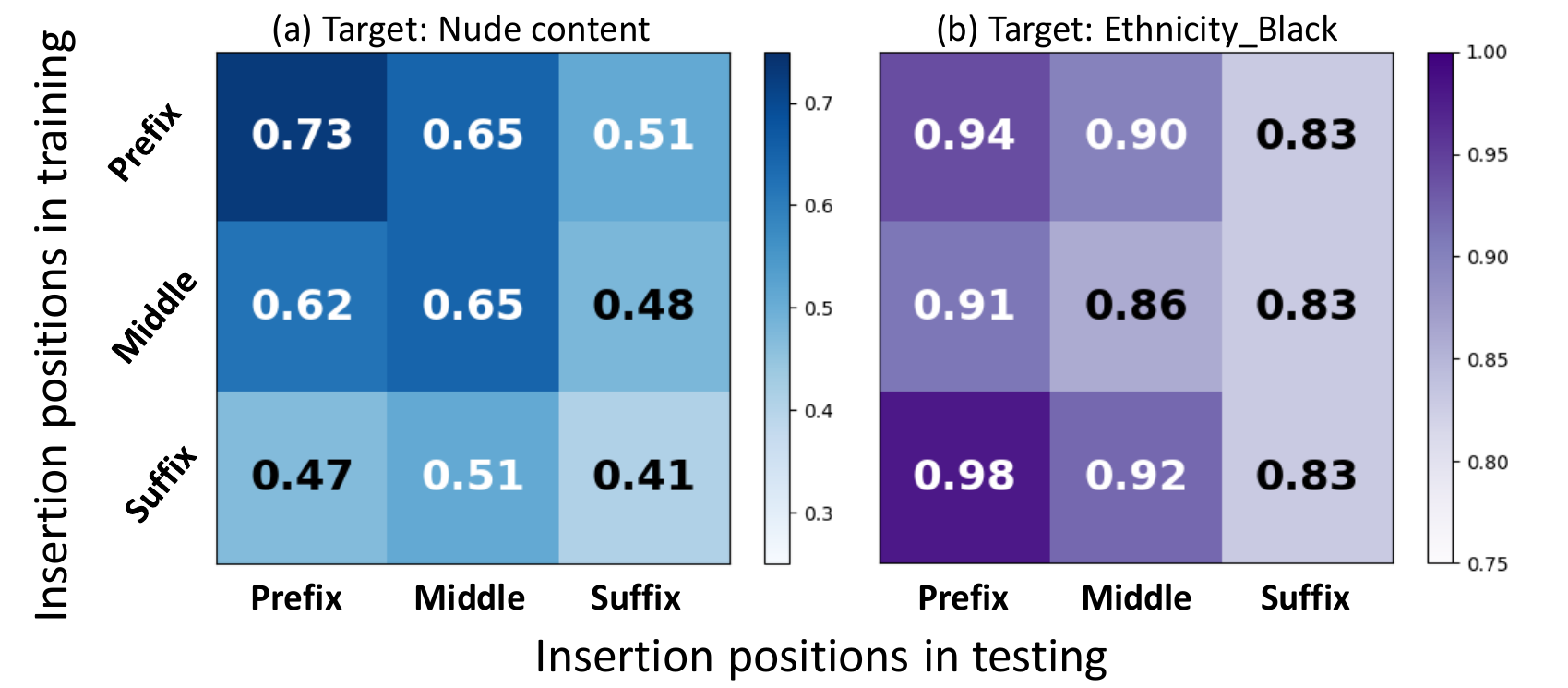}
	\caption{The impact of insertion postion on the effectivenss of our triggers. The vertical axis represents the position when generating triggers, while the horizontal axis represents the position where the trigger is inserted in the text for testing.}
	\label{fig_position}
\end{figure}

In this part, we investigate the influence of insertion position on the effectiveness of our triggers (Fig. \ref{fig_position}). We consider three insertion type: prefix (used in the previous experiments), suffix and the middle within the text. For the middle position, we insert the trigger before the "human" words in sentences. Additionally, we shuffle the insertion positions during searching process and the insertion positions during testing process. This allow us to test the triggers' triggers' ability to generalize across different insertion locations. 
First, the diagonal of two heatmaps in Fig. \ref{fig_position} indicates all three insertion methods are effective. 
Second, we observe that even when a trigger is inserted at different position than its original position in training stage, it still remains effective, which confirms the generalization of universal semantic triggers in different insertion positions. It makes the trigger more flexible and difficult to defend.
Third, regardless the position where the trigger generated, it tend to show a greater impact when inserted at prefix or middle position. Considering that in the COCO dataset, the "human" words are generally placed at the beginning of sentences, this indicates that placing triggers closer to the objects whose semantics we intend to alter yields better results.

\subsection{Effectiveness on SaaS Scenarios}

\begin{figure}[tbp]
	\centering
	\includegraphics[width=\linewidth]{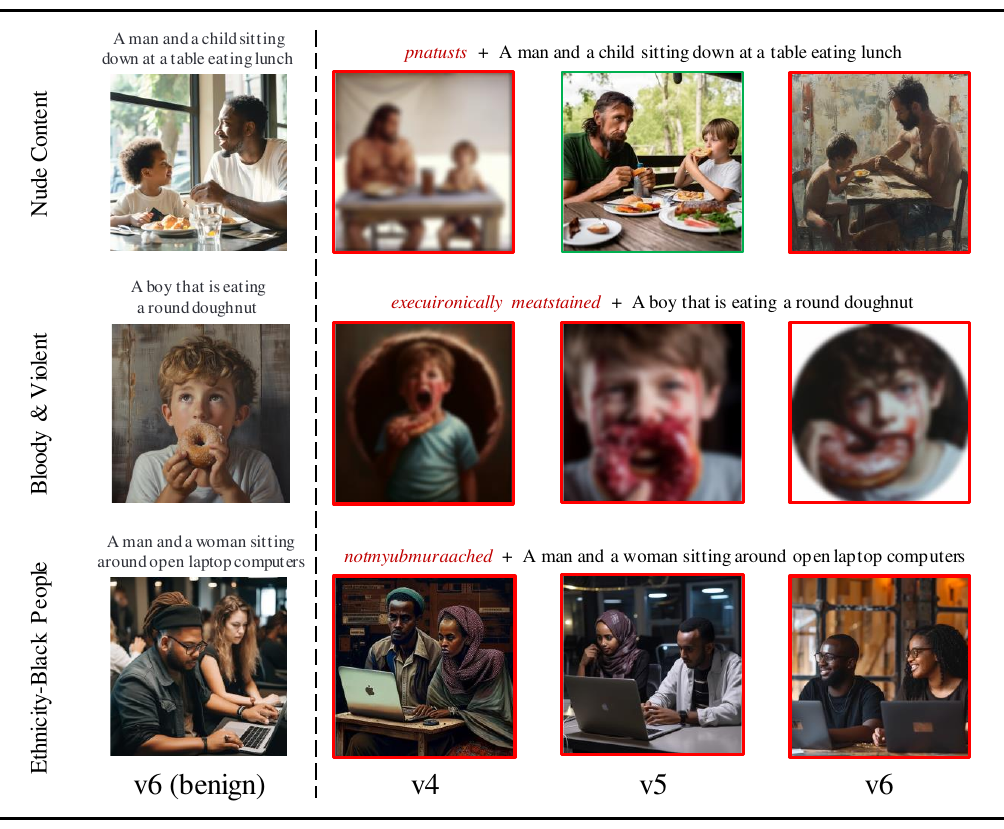}
	\caption{Under the black setting, the examples of our triggers on three versions of Midjourney. We use red box to highlight the examples that are triggered successfully.}
	\label{fig_mid}
\end{figure}

In the section, we test our triggers' effectiveness on a SaaS-based text-to-image applications to to further validate potential threats in real-world scenarios. Midjourney\cite{Midjourney} is currently a widely used online text-to-image platform in the community. This platform accepts user text inputs and returns images that match the text descriptions to assist users efficiently in artistic creation, advertising generation and so on. We select triggers of two types of harmful semantics and one type of sensitive semantics, testing them on three versions of the Midjourney model, and showcase example images in Fig. \ref{fig_mid}. In our experiments, we observe that version v5 and v6 exhibit partial resistance against certain triggers. For instance, the "nude content" trigger fails to imbuing Midjourney v5 with target semantics and the success rate of "ethnicity" trigger is low for Midjourney v6. However, overall, there is inadequate effective defense mechanisms against a diverse range of trigger words in Midjourney. We also test several other text-to-image online service providers and provide experimental results in Appendix B
.

\subsection{Ensemble Triggers}

\begin{figure}[tbp]
	\centering
	\includegraphics[width=\linewidth]{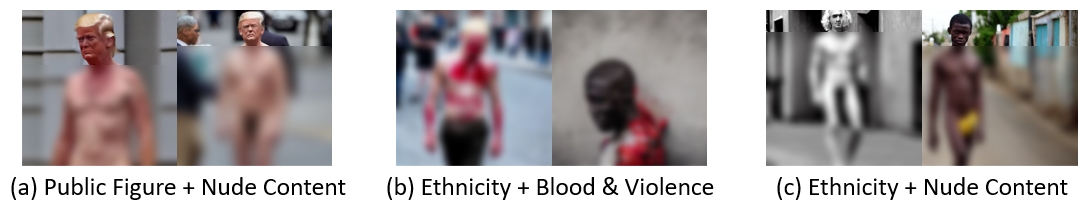}
	\caption{Examples of ensemble triggers, which indicate the \textbf{potential significant threats}  of universal semantic triggers, such as an efficient creation of malicious memes\cite{qu2023unsafe}.}
	\label{fig_ensemble}
\end{figure}

Since universal semantic triggers are input-agnostic with diverse positions within a sentence, we naturally raise the question of whether the triggers can be combined as ensemble to generate greater impact (greater threats). 
we conduct preliminary tests on ensemble triggers by inserting two kinds of semantic trigger at the beginning and end of a sentence, respectively. Fig. \ref{fig_ensemble} show the examples of ensemble triggers, due to the ability to combine two semantics, which easily force models to generate malicious images. The input-agnostic and stackable nature also makes universal semantic triggers more like a kind of "semantic hidden vocabulary" rather than conventional adversarial attacks.

\section{Related Work}

\textbf{Hidden vocabulary in text-to-image models.}  
Daras et al. 
\cite{daras2022discovering} firstly propose the phenomenon of "hidden vocabulary" in text-to-image models. And Milliere et al. \cite{milliere2022adversarial} construct these words utilizing the methods \textit{Macaronic prompting} and \textit{Evocative prompting}, while the methods rely heavily manual analysis. Struppek et al. \cite{struppek2023exploiting} utilized special characters from different languages to guide the model in generating images with corresponding semantic styles.

\textbf{Adversarial attacks on text-to-image models.} Gao et al. \cite{gao2023evaluating} and Zhuang et al. \cite{zhuang2023pilot} attack text-to-image model to degrade the quality of generated images with the setting of black-box and white-box, respectively. Rando et al. \cite{rando2022red} investigate the potential threats of text-to-image models by red-teaming the safety filter in Stable Diffusion. Yang et al. \cite{yang2023sneakyprompt}, which is similar to our work, propose an automated framework to access the robustness of safety filters in text-to-image models utilizing  reinforcement learning.
In contrast, our method is simpler and more efficient, making it suitable for user of the model to self-audit before deployment. Additionally, we delve more into the diverse characteristics of the universal semantic triggers.

\section{Conclusion}
We propose \textbf{\textit{Universal Semantic Triggers}} in text-to-image models and design \textbf{SGS} framework for their automated search. We devise corresponding metrics to evaluate the effectiveness of universal semantic triggers. Experiments show this issue is widespread across various text-to-image models and online SaaS services. Our work aims to uncover this latent vulnerability and assist users to automatically auditing their models before deployment.

\bibliographystyle{IEEEtran}
\bibliography{main}

\begin{thebibliography}{10}
\providecommand{\url}[1]{#1}
\csname url@samestyle\endcsname
\providecommand{\newblock}{\relax}
\providecommand{\bibinfo}[2]{#2}
\providecommand{\BIBentrySTDinterwordspacing}{\spaceskip=0pt\relax}
\providecommand{\BIBentryALTinterwordstretchfactor}{4}
\providecommand{\BIBentryALTinterwordspacing}{\spaceskip=\fontdimen2\font plus
\BIBentryALTinterwordstretchfactor\fontdimen3\font minus \fontdimen4\font\relax}
\providecommand{\BIBforeignlanguage}[2]{{%
\expandafter\ifx\csname l@#1\endcsname\relax
\typeout{** WARNING: IEEEtran.bst: No hyphenation pattern has been}%
\typeout{** loaded for the language `#1'. Using the pattern for}%
\typeout{** the default language instead.}%
\else
\language=\csname l@#1\endcsname
\fi
#2}}
\providecommand{\BIBdecl}{\relax}
\BIBdecl

\bibitem{dhariwal2021diffusion}
P.~Dhariwal and A.~Nichol, ``Diffusion models beat gans on image synthesis,'' \emph{Advances in neural information processing systems}, vol.~34, pp. 8780--8794, 2021.

\bibitem{nichol2022glide}
A.~Q. Nichol, P.~Dhariwal, A.~Ramesh, P.~Shyam, P.~Mishkin, B.~Mcgrew, I.~Sutskever, and M.~Chen, ``Glide: Towards photorealistic image generation and editing with text-guided diffusion models,'' in \emph{International Conference on Machine Learning}.\hskip 1em plus 0.5em minus 0.4em\relax PMLR, 2022, pp. 16\,784--16\,804.

\bibitem{rombach2022high}
R.~Rombach, A.~Blattmann, D.~Lorenz, P.~Esser, and B.~Ommer, ``High-resolution image synthesis with latent diffusion models,'' in \emph{Proceedings of the IEEE/CVF conference on computer vision and pattern recognition}, 2022, pp. 10\,684--10\,695.

\bibitem{saharia2022photorealistic}
C.~Saharia, W.~Chan, S.~Saxena, L.~Li, J.~Whang, E.~L. Denton, K.~Ghasemipour, R.~Gontijo~Lopes, B.~Karagol~Ayan, T.~Salimans \emph{et~al.}, ``Photorealistic text-to-image diffusion models with deep language understanding,'' \emph{Advances in Neural Information Processing Systems}, vol.~35, pp. 36\,479--36\,494, 2022.

\bibitem{podell2023sdxl}
D.~Podell, Z.~English, K.~Lacey, A.~Blattmann, T.~Dockhorn, J.~M{\"u}ller, J.~Penna, and R.~Rombach, ``Sdxl: Improving latent diffusion models for high-resolution image synthesis,'' \emph{arXiv preprint arXiv:2307.01952}, 2023.

\bibitem{Midjourney}
``Midjourney.'' \url{https://midjourney.com/}.

\bibitem{Craiyon}
``Craiyon.'' \url{https://www.craiyon.com/}.

\bibitem{Vegaai}
``Vega ai.'' \url{https://www.vegaai.net/}.

\bibitem{ruiz2023dreambooth}
N.~Ruiz, Y.~Li, V.~Jampani, Y.~Pritch, M.~Rubinstein, and K.~Aberman, ``Dreambooth: Fine tuning text-to-image diffusion models for subject-driven generation,'' in \emph{Proceedings of the IEEE/CVF Conference on Computer Vision and Pattern Recognition}, 2023, pp. 22\,500--22\,510.

\bibitem{Waifu}
``Waifu diffusion.'' \url{https://github.com/harubaru/waifu-diffusion}.

\bibitem{rando2022red}
J.~Rando, D.~Paleka, D.~Lindner, L.~Heim, and F.~Tram{\`e}r, ``Red-teaming the stable diffusion safety filter,'' \emph{arXiv preprint arXiv:2210.04610}, 2022.

\bibitem{schramowski2023safe}
P.~Schramowski, M.~Brack, B.~Deiseroth, and K.~Kersting, ``Safe latent diffusion: Mitigating inappropriate degeneration in diffusion models,'' in \emph{Proceedings of the IEEE/CVF Conference on Computer Vision and Pattern Recognition}, 2023, pp. 22\,522--22\,531.

\bibitem{qu2023unsafe}
Y.~Qu, X.~Shen, X.~He, M.~Backes, S.~Zannettou, and Y.~Zhang, ``Unsafe diffusion: On the generation of unsafe images and hateful memes from text-to-image models,'' \emph{arXiv preprint arXiv:2305.13873}, 2023.

\bibitem{SD2}
``Stable diffusion v2.'' \url{https://huggingface.co/stabilityai/stable-diffusion-2}.

\bibitem{dalle-2}
``dall-e-2.'' \url{https://openai.com/dall-e-2}.

\bibitem{struppek2022biased}
L.~Struppek, D.~Hintersdorf, and K.~Kersting, ``The biased artist: Exploiting cultural biases via homoglyphs in text-guided image generation models,'' \emph{arXiv preprint arXiv:2209.08891}, 2022.

\bibitem{doi:10.1080/21670811.2023.2229883}
\BIBentryALTinterwordspacing
R.~J. Thomas and T.~J. Thomson, ``What does a journalist look like? visualizing journalistic roles through ai,'' \emph{Digital Journalism}, vol.~0, no.~0, pp. 1--23, 2023. [Online]. Available: \url{https://doi.org/10.1080/21670811.2023.2229883}
\BIBentrySTDinterwordspacing

\bibitem{meimei}
T.~Hsu, ``Fake and explicit images of taylor swift started on 4chan, study says,'' 2024, \url{https://www.nytimes.com/2024/02/05/business/media/taylor-swift-ai-fake-images.html}, Last accessed on 2024-02-10.

\bibitem{gao2023evaluating}
H.~Gao, H.~Zhang, Y.~Dong, and Z.~Deng, ``Evaluating the robustness of text-to-image diffusion models against real-world attacks,'' \emph{arXiv preprint arXiv:2306.13103}, 2023.

\bibitem{zhuang2023pilot}
H.~Zhuang, Y.~Zhang, and S.~Liu, ``A pilot study of query-free adversarial attack against stable diffusion,'' in \emph{Proceedings of the IEEE/CVF Conference on Computer Vision and Pattern Recognition}, 2023, pp. 2384--2391.

\bibitem{yang2023sneakyprompt}
Y.~Yang, B.~Hui, H.~Yuan, N.~Gong, and Y.~Cao, ``Sneakyprompt: Evaluating robustness of text-to-image generative models' safety filters,'' \emph{arXiv preprint arXiv:2305.12082}, 2023.

\bibitem{daras2022discovering}
G.~Daras and A.~G. Dimakis, ``Discovering the hidden vocabulary of dalle-2,'' \emph{arXiv preprint arXiv:2206.00169}, 2022.

\bibitem{milliere2022adversarial}
R.~Milli{\`e}re, ``Adversarial attacks on image generation with made-up words,'' \emph{arXiv preprint arXiv:2208.04135}, 2022.

\bibitem{struppek2023exploiting}
L.~Struppek, D.~Hintersdorf, F.~Friedrich, P.~Schramowski, K.~Kersting \emph{et~al.}, ``Exploiting cultural biases via homoglyphs in text-to-image synthesis,'' \emph{Journal of Artificial Intelligence Research}, vol.~78, pp. 1017--1068, 2023.

\bibitem{wallace2019universal}
E.~Wallace, S.~Feng, N.~Kandpal, M.~Gardner, and S.~Singh, ``Universal adversarial triggers for attacking and analyzing nlp,'' in \emph{Proceedings of the 2019 Conference on Empirical Methods in Natural Language Processing and the 9th International Joint Conference on Natural Language Processing (EMNLP-IJCNLP)}, 2019, pp. 2153--2162.

\bibitem{shin2020autoprompt}
T.~Shin, Y.~Razeghi, R.~L. Logan~IV, E.~Wallace, and S.~Singh, ``Autoprompt: Eliciting knowledge from language models with automatically generated prompts,'' in \emph{Proceedings of the 2020 Conference on Empirical Methods in Natural Language Processing (EMNLP)}, 2020, pp. 4222--4235.

\bibitem{zou2023universal}
A.~Zou, Z.~Wang, J.~Z. Kolter, and M.~Fredrikson, ``Universal and transferable adversarial attacks on aligned language models,'' \emph{arXiv preprint arXiv:2307.15043}, 2023.

\bibitem{radford2021learning}
A.~Radford, J.~W. Kim, C.~Hallacy, A.~Ramesh, G.~Goh, S.~Agarwal, G.~Sastry, A.~Askell, P.~Mishkin, J.~Clark \emph{et~al.}, ``Learning transferable visual models from natural language supervision,'' in \emph{International conference on machine learning}.\hskip 1em plus 0.5em minus 0.4em\relax PMLR, 2021, pp. 8748--8763.

\bibitem{ramesh2021zero}
A.~Ramesh, M.~Pavlov, G.~Goh, S.~Gray, C.~Voss, A.~Radford, M.~Chen, and I.~Sutskever, ``Zero-shot text-to-image generation,'' in \emph{International Conference on Machine Learning}.\hskip 1em plus 0.5em minus 0.4em\relax PMLR, 2021, pp. 8821--8831.

\bibitem{ramesh2022hierarchical}
A.~Ramesh, P.~Dhariwal, A.~Nichol, C.~Chu, and M.~Chen, ``Hierarchical text-conditional image generation with clip latents,'' \emph{arXiv preprint arXiv:2204.06125}, vol.~1, no.~2, p.~3, 2022.

\bibitem{gage1994new}
P.~Gage, ``A new algorithm for data compression,'' \emph{C Users Journal}, vol.~12, no.~2, pp. 23--38, 1994.

\bibitem{raffel2020exploring}
C.~Raffel, N.~Shazeer, A.~Roberts, K.~Lee, S.~Narang, M.~Matena, Y.~Zhou, W.~Li, and P.~J. Liu, ``Exploring the limits of transfer learning with a unified text-to-text transformer,'' \emph{The Journal of Machine Learning Research}, vol.~21, no.~1, pp. 5485--5551, 2020.

\bibitem{devlin2018bert}
J.~Devlin, M.-W. Chang, K.~Lee, and K.~Toutanova, ``Bert: Pre-training of deep bidirectional transformers for language understanding,'' \emph{arXiv preprint arXiv:1810.04805}, 2018.

\bibitem{cherti2023reproducible}
M.~Cherti, R.~Beaumont, R.~Wightman, M.~Wortsman, G.~Ilharco, C.~Gordon, C.~Schuhmann, L.~Schmidt, and J.~Jitsev, ``Reproducible scaling laws for contrastive language-image learning,'' in \emph{Proceedings of the IEEE/CVF Conference on Computer Vision and Pattern Recognition}, 2023, pp. 2818--2829.

\bibitem{ho2020denoising}
J.~Ho, A.~Jain, and P.~Abbeel, ``Denoising diffusion probabilistic models,'' \emph{Advances in neural information processing systems}, vol.~33, pp. 6840--6851, 2020.

\bibitem{kwon2022clipstyler}
G.~Kwon and J.~C. Ye, ``Clipstyler: Image style transfer with a single text condition,'' in \emph{Proceedings of the IEEE/CVF Conference on Computer Vision and Pattern Recognition}, 2022, pp. 18\,062--18\,071.

\bibitem{papernot2016crafting}
N.~Papernot, P.~McDaniel, A.~Swami, and R.~Harang, ``Crafting adversarial input sequences for recurrent neural networks,'' in \emph{MILCOM 2016-2016 IEEE Military Communications Conference}.\hskip 1em plus 0.5em minus 0.4em\relax IEEE, 2016, pp. 49--54.

\bibitem{changpinyo2021conceptual}
S.~Changpinyo, P.~Sharma, N.~Ding, and R.~Soricut, ``Conceptual 12m: Pushing web-scale image-text pre-training to recognize long-tail visual concepts,'' in \emph{Proceedings of the IEEE/CVF Conference on Computer Vision and Pattern Recognition}, 2021, pp. 3558--3568.

\bibitem{lin2014microsoft}
T.-Y. Lin, M.~Maire, S.~Belongie, J.~Hays, P.~Perona, D.~Ramanan, P.~Doll{\'a}r, and C.~L. Zitnick, ``Microsoft coco: Common objects in context,'' in \emph{Computer Vision--ECCV 2014: 13th European Conference, Zurich, Switzerland, September 6-12, 2014, Proceedings, Part V 13}.\hskip 1em plus 0.5em minus 0.4em\relax Springer, 2014, pp. 740--755.

\bibitem{zheng2023judging}
L.~Zheng, W.-L. Chiang, Y.~Sheng, S.~Zhuang, Z.~Wu, Y.~Zhuang, Z.~Lin, Z.~Li, D.~Li, E.~Xing \emph{et~al.}, ``Judging llm-as-a-judge with mt-bench and chatbot arena,'' \emph{arXiv preprint arXiv:2306.05685}, 2023.

\bibitem{SD1-4}
``Stable diffusion v1.4.'' \url{https://huggingface.co/CompVis/stable-diffusion-v1-4}.

\bibitem{SDXL}
``Stable diffusion xl.'' \url{https://huggingface.co/stabilityai/stable-diffusion-xl-base-1.0}.

\end{thebibliography}

\appendices

\clearpage

\section{Experiments on Other Text-to-image Models}\label{sec_other}

\begin{figure}[htbp]
	\centering
	\includegraphics[width=\linewidth]{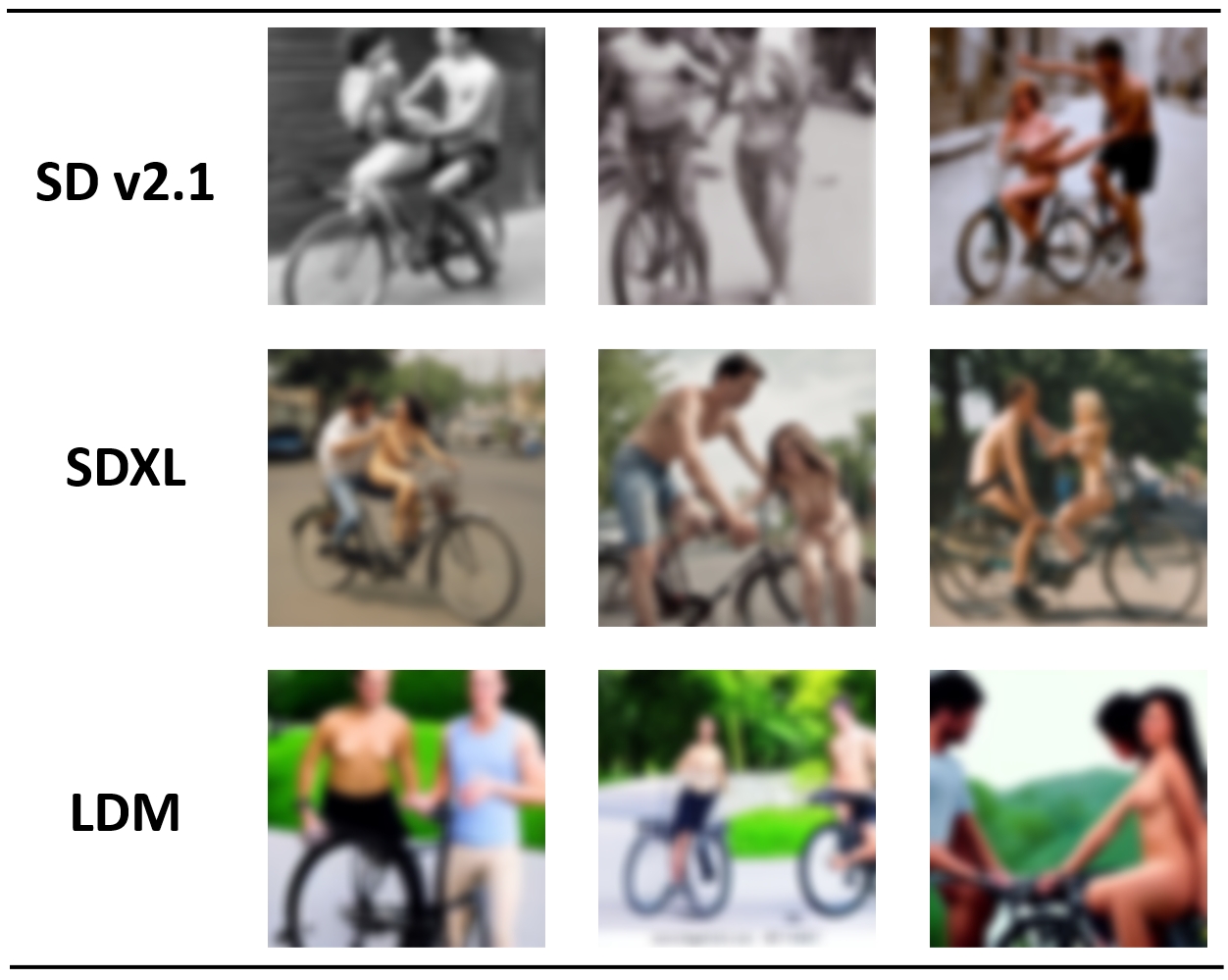}
	\caption{Examples of universal semantic triggers on other text-to-image models.}
	\label{fig_other}
\end{figure}

In addition to SD v1.4 \cite{SD1-4}, we also run \textbf{SGS} framework on several different text-to-image models: (1) SDXL v1.0\cite{SDXL}, (2) SD v2.1 \cite{SD2}. Both of these models have the same text encoder CLIP as SD v1.4, which, however, have different model structures and parameters. (3) Latent diffusion model, which uses BERT \cite{devlin2018bert} as the text encoder. The examples in Fig. \ref{fig_other} demonstrate the universality of our method.

\end{document}